\documentclass[prl,twocolumn,amsmath,amssymb,superscriptaddress]{revtex4}
\usepackage{graphicx}
\usepackage{color}
\begin{document}

\title{Designing hyperchaos and intermittency in semiconductor superlattices}

\author{E. Momp\'o}

\affiliation{Gregorio Mill\'an Institute for Fluid Dynamics, Nanoscience
and Industrial Mathematics, and Department of Mathematics, Universidad Carlos III de Madrid, 28911 Legan\'{e}s, Spain}
\author{ M. Carretero}
\affiliation{Gregorio Mill\'an Institute for Fluid Dynamics, Nanoscience
and Industrial Mathematics, and Department of Mathematics, Universidad Carlos III de Madrid, 28911 Legan\'{e}s, Spain}

\author{L. L. Bonilla}
\affiliation{Gregorio Mill\'an Institute for Fluid Dynamics, Nanoscience
and Industrial Mathematics, and Department of Mathematics, Universidad Carlos III de Madrid, 28911 Legan\'{e}s, Spain}\affiliation{Corresponding author. E-mail: bonilla@ing.uc3m.es}

\begin{abstract}
Weakly coupled semiconductor superlattices under dc voltage bias are excitable systems with many degrees of freedom that may exhibit spontaneous chaos at room temperature and act as fast physical random number generator devices. Superlattices with identical periods exhibit current self-oscillations due to the dynamics of charge dipole waves but chaotic oscillations exist on narrow voltage intervals. They disappear easily due to variation in structural growth parameters. Based on numerical simulations, we predict that inserting two identical sufficiently separated wider wells increases superlattice excitability by allowing wave nucleation at the modified wells and more complex dynamics. This system exhibits hyperchaos and varieties of intermittent chaos in extended dc voltage ranges. Unlike in ideal superlattices, our chaotic attractors are robust and resilient against noises and against controlled random disorder due to growth fluctuations. 
\end{abstract} 

\bigskip

\noindent {\it Date}: {\today}\\


\maketitle

\renewcommand{\thefootnote}{\arabic{footnote}}
\newcommand{\fin}{\newline \rule{2mm}{2mm}}

\paragraph{Introduction.} 
Recently, spontaneous chaotic oscillations in semiconductor superlattices (SSLs) at room temperature \cite{hua12} have attracted attention as all-electronic fast generators of true random numbers \cite{li13}, which are crucial to secure fast  safe data storage and transmission \cite{sti95,gal08,nie00}, stochastic modeling \cite{asm07}, and Monte Carlo simulations \cite{bin02}. Quantum partition noise \cite{bla00,rey03,fev18} due to electron tunneling is the origin of randomness at the bottom of entropy generation whereas most of the unpredictability of the final number sequence rests on chaotic evolution: random quantum seeds are expanded into fast changing physical processes achieving generation rates up to hundreds of Gb/s \cite{li13}. These superlattice devices are smaller and more scalable than similarly fast optoelectronic devices based on chaotic semiconductor lasers \cite{uch08,rei09,kan10,sci15,tan15}. SSLs are nonlinear systems with many degrees of freedom, whose effective nonlinearity originates from the well-to-well sequential resonant tunneling process \cite{kas95,bon02r,bon05,BT10}. Very different spatiotemporal patterns observed in dc-biased SSLs include static high-field domains, excitability due to collective charge dynamics, and self-sustained periodic and quasiperiodic current oscillations at low temperatures \cite{kas95,bon02r,bon05,BT10}. Similar nonlinear excitable systems include lithium batteries \cite{dre10}, modular proteins, DNA hairpins \cite{bon15}, and peristaltic fluid motion in slime molds \cite{ali13}. Oscillatory and chaotic phenomena alike those in SSLs have been predicted and observed in quantum cascade lasers \cite{alm19,spi19,ond21}. External noise affects remarkably SSLs. It induces current switching with nonexponential kinetics \cite{bom12}. At room temperature, large amplitude noise having much smaller bandwidth than the oscillation frequency creates/enhances chaotic oscillations over wider voltage ranges \cite{yin17}. Chaos synchronization between SSLs has been demonstrated \cite{li15}. Sufficiently strong external noise induces self-sustained oscillations in an otherwise stationary state (coherence resonance) and helps detecting superimposed weak periodic signals by stochastic resonance \cite{mom18,sha18}.

Achieving better control of SSL random number generators (RNGs) requires understanding better spontaneous chaos at room temperature. Current theoretical knowledge is based on numerically simulating sequential tunneling electron transport models for {\em ideal}\, SSLs having  identical periods. Spontaneous chaos is very sensitive to voltage bias, is enhanced by noise \cite{alv14,bon17,yin17}, and depends strongly on SSL configurations: short SSLs display clear period doubling cascades to chaos, which occur on narrower voltage intervals for longer SSLs \cite{rui17,ess18}. See also \cite{ama02,ama03} for period doubling routes to chaos in ideal SSLs at low temperatures. Random imperfections strongly affect chaos and can alter significantly these scenarios  \cite{ess18}. Overall, numerical simulations predict spontaneous chaos on narrower voltage intervals than those reported in experiments, which means we need  better modeling. In contrast to this situation, theoretical predictions of {\em driven chaos} under ac+dc voltage bias based on a simple sequential tunneling model \cite{bul95} are robust and were observed in experiments many years ago \cite{luo98}.

Here, we adopt a different point of view: Can we {\em design imperfections} in SSLs to produce spontaneous chaos? The answer is yes. We show by numerical simulating a sequential tunneling model of nonideal SSLs that inserting a wider well in a long SSL may trigger dipole waves there. Our design idea is that appropriately modifying one or more wells in the SSL triggers randomly dipole waves, yielding hyperchaos, chaos with more than one positive Lyapunov exponent. It transpires that well modifications and their location are crucial to produce rich chaotic dynamics. Another important question is how resilient is hyperchaos to fluctuations and additional imperfections. We investigate this point to show that our design of strong chaotic attractors is robust.

\paragraph{Microscopic sequential tunneling model.} 
Commonly used models of SSL electron transport based on sequential tunneling are reviewed in \cite{bon02r,wac02,bon05,BT10,bon17}. In these models, each SSL period is described by average values of the electric field and electron density. The effective masses and permittivities of the different materials comprising the SSL are replaced by average values. Here, we treat barriers and wells as separated entities \cite{agu97,bon00}, which describes  more completely and realistically electron transport and spontaneous chaos. In weakly coupled SSLs, intrasubband scattering times are much shorter than intersubband scattering times which, in turn, are much shorter than the interwell tunneling time across barriers. In processes varying on the latter scale, only the lower subband $C1$ is occupied, electrons are in local Fermi-Dirac equilibrium within it, and their two dimensional (2D) electron density $n_i$ is related to their chemical potential $\mu_i$ by \cite{bon02r}
\begin{eqnarray}
n_i = \frac{m_Wk_BT}{\pi\hbar^2}\int_0^\infty A_{C1_i}(\epsilon)\, \ln\!\left(1+e^{(\mu_i-\epsilon)/k_BT}\right) d\epsilon.   \label{eq1}
\end{eqnarray}
Here $i=1,\ldots, N$ ($N$ is the number of SSL periods); $d_{W_i}$ and $d_{B_i}$ are the widths of the $i$th well and barrier, respectively. The electron temperature equals the lattice temperature $T$; $k_B$ is the Boltzmann constant. $m_W$ and $m_{B_i}$ are the electron effective mass at the GaAs wells and at the barrier $i$, respectively. The Lorentzian functions $A_{C1_i}(\epsilon) = (\gamma_1/\pi)/[(\epsilon - \mathcal{E}_{C1_i})^{2} +\gamma_1^{2}]$ account for the widening of the C1 subband energies $\mathcal{E}_{C1_i}$ due to scattering with lifetime $\tau_{sc}=\hbar/\gamma_1$ 
\cite{bon00}. 

Poisson equations relate the voltage drops in the barriers, $V_{i}$, and the wells, $V_{w_i}$:
\begin{subequations}\label{eq3}
\begin{eqnarray}
\varepsilon_{W}\,\frac{V_{w_{i}}}{d_{W_i}} &=&\varepsilon_{B_{i-1}}\,\frac{V_{i-1}}{d_{B_{i-1}}}+\frac{e}{2}\left(n_i-N_{D_i}\right)\!, \label{eq3a}\\
\varepsilon_{B_{i}}\frac{V_{i}}{d_{B_{i}}}&=&\varepsilon_{B_{i-1}}\frac{V_{i-1}}{d_{B_{i-1}}} +e\left(n_i-N_{D_i}\right)\!,  \label{eq3b}
\end{eqnarray}
\end{subequations}
where $-e<0$, $\varepsilon_W$ and $\varepsilon_{B_i}$ and $N_{Di}$ are the electron charge, well and barrier static permittivities and the 2D intentional doping density at the well $i$, respectively \cite{gol87a,bon00}. The barrier $i=0$ separates the injector region from the SSL proper. Eqs.~(\ref{eq3}) imply
$V_{w_i}= \frac{d_{W_i}}{2\varepsilon_{W}}\left(\frac{\varepsilon_{B_{i-1}}V_{i-1}}{d_{B_{i-1}}} + \frac{\varepsilon_{B_i}V_{i}}{d_{B_i}}\right)\!,$ 
thereby eliminating the potential drops at the wells from the system \cite{bon00}.

The voltage drops at the barriers satisfy Amp\`ere's law
\begin{eqnarray}
 \frac{\varepsilon_{B_i}}{d_{B_i}}\frac{dV_{i}}{dt} + J_{i\to i+1} +\xi_i(t) = J(t), \label{eq5}
 \end{eqnarray}
where $J(t)$ is the total current density, $J_{i\to i+1}=J_{i\to i+1}(V_{i-1},V_i,V_{i+1},\mu_i,\mu_{i+1})$ is the tunneling current through barrier $i$, and $\xi_i(t)$ the corresponding fluctuating current \cite{bon17}. See  \cite{suppl} for explicit formulas. Time differentiating \eqref{eq3b} and using \eqref{eq5} produce a spatially discrete continuity equation for the electron density $n_i$. Boundary conditions at emitter and collector are phenomenological Ohm laws, $J_{0\to 1}= \sigma_e\frac{V_0}{d_{B_0}}$, and  $J_{N\to N+1}=\sigma_c \frac{n_N}{N_{D_N}}\,\frac{V_N}{d_{B_N}}$, where $\sigma_e$ and $\sigma_c$ are the contact conductivities, $d_{B_j}$ are effective lengths for the contact regions and $N_{D_N}$ is the collector effective 2D doping density, cf. \cite{ama01}. The voltage bias condition $V= \sum_{i=0}^{N}V_{i}+\sum_{i=1}^{N}V_{w_i}=V_{dc}+\eta(t)$ closes the set of equations; $\eta(t)$ is the external circuit noise. 

\paragraph{Modified SSL and deterministic numerical simulations.}
Refs.~\cite{li13,hua12} consider $50$-period GaAs/Al$_{0.45}$Ga$_{0.55}$As SSLs with $d_W=7$ nm, $d_B=4$ nm. Wells have three subbands with energies 41.6, 165.8, and 354.3 meV and level broadenings due to scattering, 2.5, 8 and 24 meV, respectively \cite{bon06}. Effective electron mass and permittivities at wells and barriers are $m_W=0.063\, m_e$, $m_B=(0.063 + 0.083x)m_e=0.1 m_e$, $\varepsilon_B= 10.9 \epsilon_0$, $\varepsilon_W= (12.9- 2.84x) \epsilon_0=11.7\epsilon_0$ ($x=0.45$; $\epsilon_0$ is the vacuum dielectric constant), respectively. The central part of all quantum wells is doped, producing an equivalent 2D doping density $N_D=6 \times 10^{10}$ cm$^{-2}$. $A=s^2$, $s=30\,\mu$m, $l=d_B+d_W$, are the SSL cross section, the square mesa side length, and the SSL period, respectively. $\sigma_c=\sigma_e= 0.49$ A/(Vm),   $N_{D_N}=N_D$. 

Firstly, ignore noises and fluctuations in doping density and in barrier and well widths. The $I-V$ characteristics of  SSLs with identical periods correspond to stable stationary states except on voltage intervals where self-sustained oscillations of the current exist (provided the number of SSL periods exceeds 14, the minimum number needed to attain self-oscillations \cite{BT10}). Time periodic self-oscillations are caused by the formation of dipole waves at the emitter, motion toward, and annihilation at the collector. For the numerical parameter values listed above and increasing voltage, the self-oscillation branch starts at a supercritical Hopf bifurcation and ends at a saddle-node infinite period (SNIPER) bifurcation \cite{mom18,sha18}; see Fig. 1 in \cite{suppl}. Changing parameters, we may find narrow regions of chaotic attractors \cite{alv14}, but we prefer to modify the SSL design searching robust chaos. 

\begin{figure}[htbp]
\begin{center}
\includegraphics[width=8cm,angle=0]{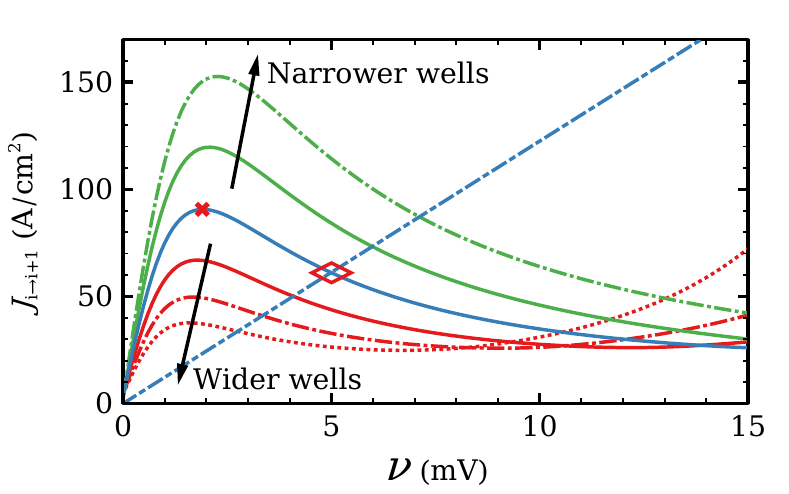}
\end{center}
\caption{Tunneling current-voltage characteristics for ideal SSLs having identical periods with $n_i=N_D$, $V_i=\mathcal{V}$. We compare the reference configuration $d_B=4$ nm, $d_W=7$ nm to the contact Ohm's law and to other configurations with more or less monolayers at all the wells. Wider (narrower) wells produce curves with lower (higher) maximum current. The maximum and critical currents of the reference configuration are marked with a cross and a diamond, respectively. Other configurations differ in three, six or nine monolayers from reference. Energy levels calculated using the Kronig-Penney model.} \label{iv} 
\end{figure}

What can we expect by changing the width of a well in an otherwise ideal SSL? Consider the ideal SSL tunneling current, $J_{i\to i+1}(\mathcal{V})=J_{i\to i+1}(\mathcal{V},\mathcal{V},\mathcal{V},\mu_D,\mu_D)$, for constant barrier voltage drops $V_i=\mathcal{V}$, fixed electron densities $n_i=N_D$, and $\mu_D$ given by \eqref{eq1}. Fig.~\ref{iv} shows how the intersection of $J_{i\to i+1}(\mathcal{V})$ and $J_{0\to 1}(\mathcal{V})$ (marked with a diamond for the reference configuration) changes with well width. This intersection roughly marks the {\em critical} voltage and current at which the contact issues a dipole wave, which causes current self-oscillations, excitability and other phenomena \cite{bon02r,wac02,bon05,BT10,mom18}. Clearly the critical current is lower for SSL with wider wells, which facilitates   triggering dipole waves there. The opposite holds for narrower wells. What happens when we make a single well of a long SSL wider? Numerical simulations show that each added monolayer shifts significantly the region of self-oscillations until there are 6 extra monolayers in total. From that point on, adding more monolayers to the modified well does not change the self-oscillation region of the $I-V$ curve. Thus, we insert a single well having $d_W=10$ nm (ten added monolayers to a 7-nm well) and  energy levels $\mathcal{E}_{C_1}=24.0$ meV, $\mathcal{E}_{C_2}=96.1$ meV, $\mathcal{E}_{C_3} = 214.7$ meV. Numerical simulations show $I-V$ curves with one or more regions of self-oscillations depending on the wider well position. As for unmodified SSLs, current self-oscillations arise from the dynamics of charge dipole waves, which may be nucleated at the wider well. Dipole waves change slightly when traveling through the wider well, which produces sudden and short-lived current spikes. We do not find chaos. 

\begin{figure}[htbp]
\begin{center}
\includegraphics[width=8cm,angle=0]{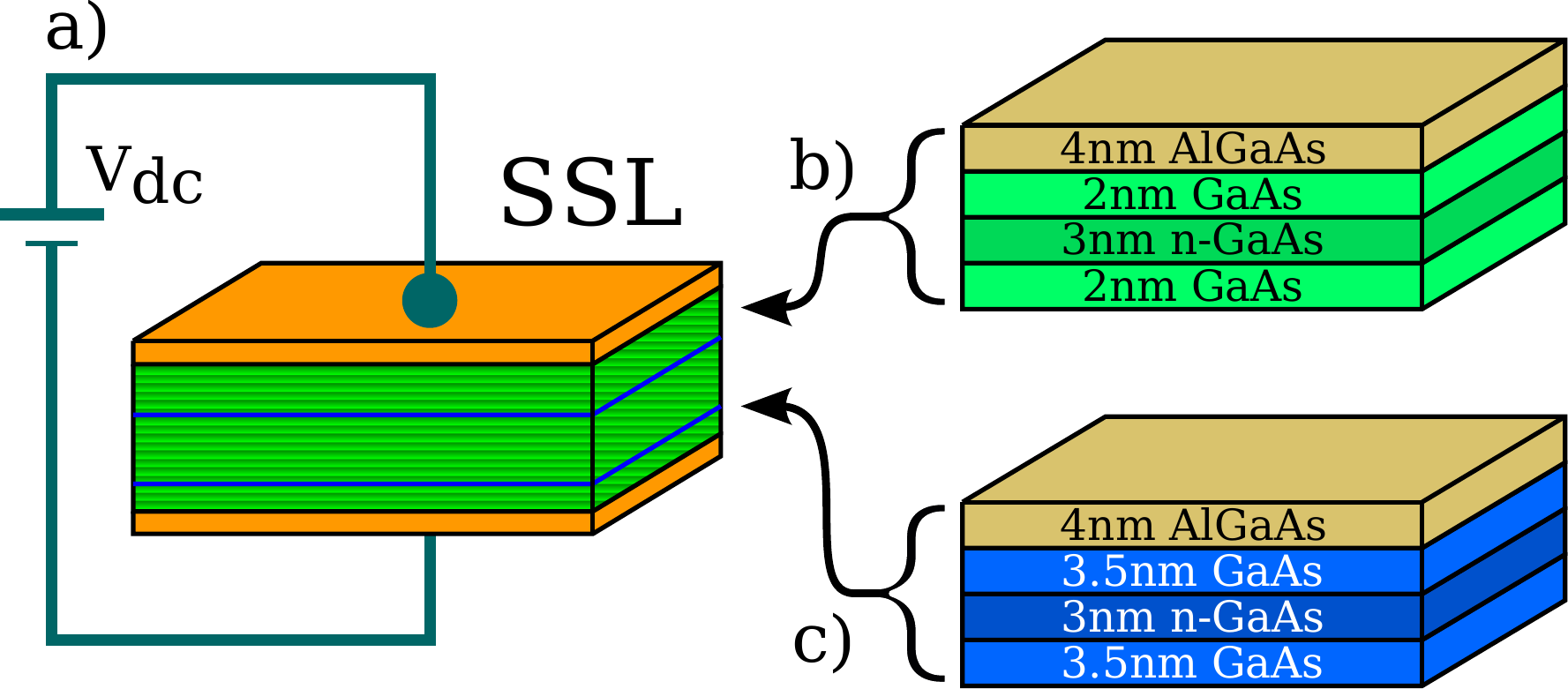}
\end{center}
\caption{(a) Sketch of circuit containing the modified SSL; (b) barrier/well period of the unmodified structure \cite{li13}; (c) barrier and modified well.} \label{sketch} 
\end{figure}

When we introduce two well separated wider wells, the resulting dynamics is dominated by one of them (and it is not chaotic) unless both modified wells are identical, as in Fig.~\ref{sketch}. The wider wells are not adjacent to the contact regions, we label them $i_1$ and $i_2$ ($i_1<i_2$), with widths $d_{W_i}=10$ nm, and regions I, II and III are the intervals $i<i_1$, $i_1<i<i_2$, and $i>i_2$, respectively. If regions II and III have more than 14 wells (minimal length to sustain oscillations \cite{BT10}), dipole waves can be nucleated at $i_1$ and at $i_2$, they travel through regions II and III respectively, and their motion is strongly correlated. Typically, each region II and III supports one dipole wave but there are specific instances of two waves moving on the same region. For $i_1=5$, observation of chaotic attractors requires the second wider well to satisfy $28< i_2\leq 35$. Note that modified SSLs exhibit self-oscillations with faster frequencies than in ideal SSLs because the dipole waves causing them travel on shorter regions of the device.

\begin{figure}[htbp]
\begin{center}
\includegraphics[width=8cm,angle=0]{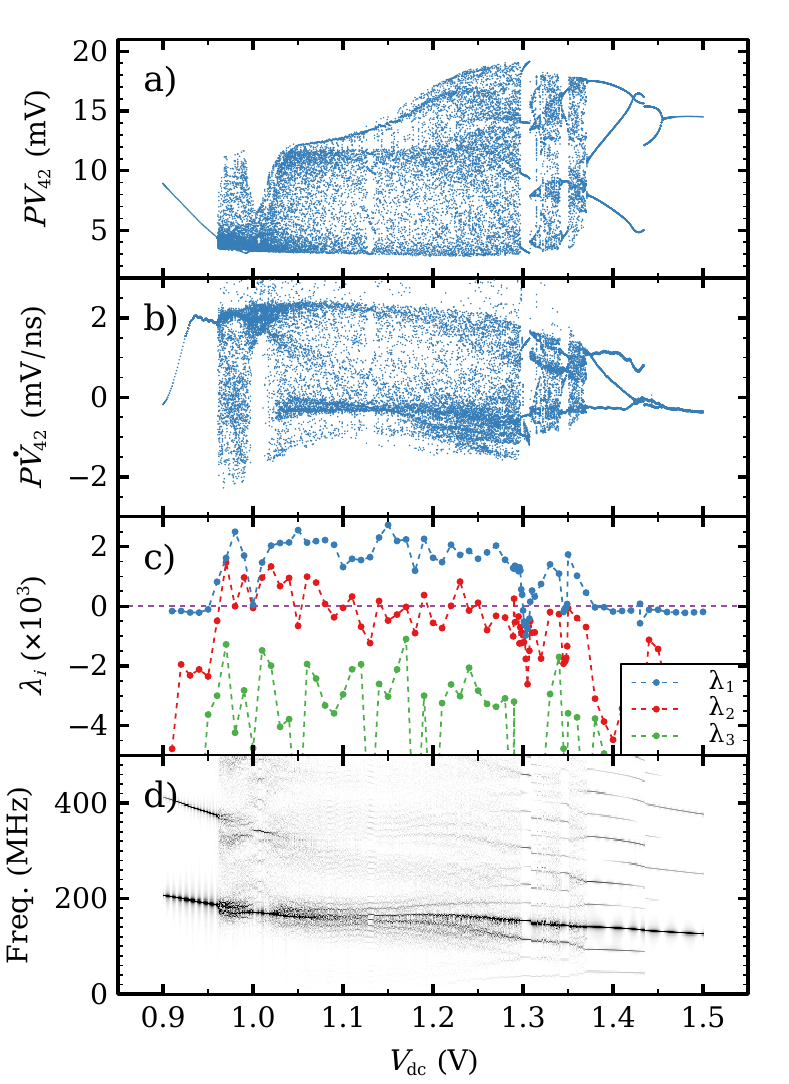}
\end{center}
\caption{(a) Poincar\'e map from $V_{42}(t)$, (b) Poincar\'e map from $\dot{V}_{42}(t)$, (c) three largest Lyapunov exponents, and (d) Fourier spectrum density plot versus $V_{dc}$ for the modified SSL with $i_1=5$ and $i_2=30$. To avoid redundant symmetric points, the Poincar\'e maps depict $V_{42}(t)$ and $\dot{V}_{42}(t)$ at times $t^*$ where the time trace at a widely separated well, $V_{12}(t)$, takes on its mean value in time and $\dot{V}_{12}(t^*)>0$. Each panel shows features hidden in the other ones. The Poincar\'e map reveals jumps between periodic attractors at $V_{dc}=1.3\text{V}$ and $V_{dc}=1.43\text{V}$. The Fourier spectrum reveals underlying quasi-periodic behavior with different incommensurate frequencies, whereas the Lyapunov exponents show that the system is hyperchaotic for $V_{dc}<1.08\text{V}$ ($\lambda_1, \lambda_2>0$ and of comparable scales). For $V_{dc}>1.08\text{V}$, the system has intermittent chaos at different time-scales ($\lambda_1\gg\lambda_2\approx 0$).} \label{fig4} 
\end{figure}

\begin{figure}[htbp]
\begin{center}
\includegraphics[width=8cm,angle=0]{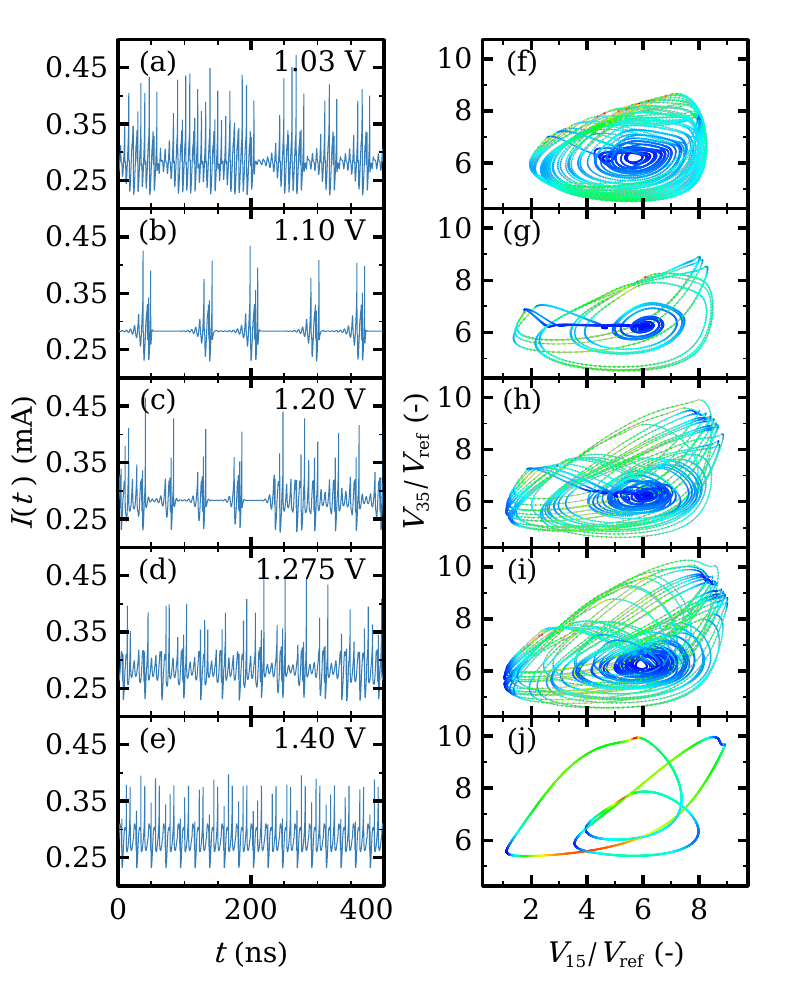}
\end{center}
\caption{(a)-(e) Current traces $I(t)$, and (f)-(j) phase plane portraits $(V_{15},V_{35})$ illustrating hyperchaos and intermittencies for $V_{dc}=$ (a,f) 1.03 V, (b,g) 1.10 V, (c,h) 1.20 V, (d,i) 1.275 V, (e,j) 1.40 V. In the phase portraits blue to red colors indicate lower to higher values of $\sqrt{\dot{V}_{15}^2+\dot{V}_{35}^2}$, thereby visualizaling rate of change in voltage drops. See also density plots of the electric field in \cite{suppl}.} \label{fig5} 
\end{figure}

Through Poincar\'e maps, Lyapunov exponents and density plots of the Fourier spectrum build from potential drops at separated periods, $V_{12}$ and $V_{42}$, Fig.~\ref{fig4} shows a variety of dynamical behaviors for the voltage range where self-oscillations occur. 
As the voltage bias $V_\text{dc}$ increases, the stationary state loses its stability and a time periodic attractor (cycle) appears at $V_\text{dc}=0.9$ V. The voltage profiles consist of charge dipole waves being repeatedly nucleated at both modified wells and advancing towards the collector without reaching it. At $V_\text{dc}=0.96$ V a second cycle appears and interacts with the first one. The result is a hyperchaotic attractor with two positive Lyapunov exponents. Trajectories fill the space between the two cycles, cf. phase portrait $(V_{15},V_{35})$ in Fig.~\ref{fig5}(f). In the voltage interval $0.96 < V_\text{dc} < 1.1$ for hyperchaos, dipole waves nucleated at the second modified well either cannot reach the collector or, if they do, dipoles cannot stay in the wells near the collector. For larger voltages, the second Lyapunov exponent becomes smaller albeit positive, and intermittent chaos appears instead, cf. Fig.~\ref{fig4}(c). This corresponds to the appearance of another cycle that interacts with the others and eventually disappears at a saddle point; see Fig.~\ref{fig5}(f) and \ref{fig5}(g) for $V_\text{dc}=1.03$ V and 1.10 V, respectively. Intermittency chaos appears for the interval $1.10 < V_\text{dc} < 1.37$: irregular bursts corresponding to a cycle are separated by intervals for which the trajectories are close to the saddle point, cf. Fig.~\ref{fig5}(c). This behavior is associated to dipole waves that reach the collector, stop there and remain in the last SSL periods (quiescent stage), whereas periodic bursts are associated to dipole wave recycling in Regions II and III \cite{suppl}. At $V_\text{dc}=1.2$ V, the saddle point expands to a saddle cycle and the intermittent behavior continues, cf. Fig.~\ref{fig5}(d). The quiescent stage is now associated to low frequency oscillations. At $V_\text{dc}=1.37$ V, the intermittency becomes a period 3 cycle: three loop trajectories in the phase plane, cf. Fig.~\ref{fig5}(j). At larger {\em dc} voltages the periodic behavior continues and it becomes simpler (two loops at 1.43 V, a single loop for larger voltages). The transition from periodic attractors with three loops to two loop ones at 1.43 V is rather abrupt; see Figs.~\ref{fig4}(a) and \ref{fig4}(b). There is a hysteresis cycle about this voltage value that becomes manifest by sweeping up or down the {\em dc} voltage. The self-oscillation last branch disappears at a supercritical Hopf bifurcation.

\paragraph{Effects of width randomness and noise.}
When growing SSLs, it is difficult to control perfectly the layer width of the two semiconductors. How do well width  fluctuations affect SSL current self-oscillations? Let $d_{W_i}+\delta_i$ be the widths, with $d_{W_5}=d_{W_{30}}=10$ nm and $d_{W_i}=7$ nm for $i\neq 5,30$. We extract $\delta_i$ out of a zero mean normal distribution with standard deviation $\sigma$. Then deviations larger than $\pm 2\sigma$ are rare. We ignore internal, external noises and  fluctuations in  doping density, barrier width and composition. Depending on the resulting random configuration, intervals of hyperchaos or intermittent chaos may be destroyed. The success rate of disordered SSLs still exhibiting chaotic behavior depends on $\sigma$, cf. Fig.~5 in \cite{suppl}. For $\sigma<0.015$ nm, the chaotic attractors of the SSL without disorder persist and $\sigma= 0.024$ nm is sufficient to yield a 70\% success rate. During epitaxial growth \cite{gra95}, Al atoms within each interface alloy monolayer may be segregated into local clusters or not be randomly positioned in the Ga or the As sublattice \cite{das12}. This produces nonzero $\sigma$ even if there are no errors in the number of monolayers per barrier and well (monolayer width: 0.3 nm). Careful design achieves $\sigma<0.018$ nm in simpler devices \cite{das12,mis15}, which would yield reliably chaotic SSLs.

In repeated simulations of SSL equations with internal (shot and thermal) and external (2 mV rms for a 50 Ohm resistor) noise, we observe that noises diminish the largest Lyapunov exponent of chaotic attractors and augment slightly the third Lyapunov exponent (which does not become positive). Noise forces the system to visit more often  contraction regions of the phase space such as the quiescent regions between bursts in intermittent chaos. This lowers the largest Lyapunov exponent \cite{zho02}. Thus, contrary to the effect reported and observed in ideal SSL with identical periods \cite{alv14,yin17}, noises do not enhance chaos in our modified SSL. 

In {\em conclusion}, inserting two wider wells (separated by at least 14 wells and not adjacent to the contacts) in a long SSL produces robust resilient chaos on wide bias ranges: hyperchaos and intermittent chaos persist in the presence of disorder and noise. Our design  is based on the complex dipole wave dynamics triggered at the two identical wider wells. It requires careful control of sample growth. Unlike our proposed structures, ideal SSLs with identical periods loose more easily their complex dynamics to well width fluctuations. Our work 
ushers in robust design of chaotic attractors in SSLs, which act as all-electronic building blocks of fast true RNGs. We could explore enhancing complex dynamics and hyperchaotic attractors by inserting more separated identical wider wells within the limits set by the minimum separation distance for self-oscillations and the longest working SSL that is feasible to build. Another important application is building distribution devices for encryption keys that exploit chaos synchronization \cite{keu17}. Since synchronization of chaotic SSLs has been experimentally demonstrated \cite{li15,liu18}, our robust design of chaotic SSL devices may be used to distribute secret keys safely. 

 \acknowledgments
We acknowledge support by the FEDER/Ministerio de Ciencia, Innovaci\'on y Universidades -- Agencia Estatal de Investigaci\'on grants MTM2017-84446-C2-2-R and PID2020-112796RB-C22, by the Madrid Government (Comunidad de Madrid-Spain) under the Multiannual Agreement with UC3M in the line of Excellence of University Professors (EPUC3M23), and in the context of the V PRICIT (Regional Programme of Research and Technological Innovation). 


\begin{thebibliography}{9}
\bibitem{hua12} Y. Y. Huang, W. Li, W. Q. Ma, H. Qin, Y. H. Zhang, Experimental observation of spontaneous chaotic current oscillations in GaAs/Al$_{0.45}$Ga$_{0.55}$As superlattices at room temperature. 
Chinese Sci. Bull. {\bf 57}, 2070 
 (2012).
\bibitem{li13} W. Li, I. Reidler, Y. Aviad, Y. Huang, H. Song, Y. Zhang, M. Rosenbluth, I. Kanter, Fast Physical Random-Number Generation Based on Room-Temperature Chaotic Oscillations in Weakly Coupled Superlattices.
Phys. Rev. Lett. {\bf 111}, 044102 (2013).
\bibitem{sti95} D. R. Stinson, Cryptography: Theory and Practice
(CRC Press, Boca Raton, 1995), The CRC Press series on discrete mathematics and its applications.
\bibitem{gal08} R. G. Gallager, Principles of Digital Communication (Cambridge University Press, Cambridge, UK, 2008).
\bibitem{nie00} M. A. Nielsen, I. L. Chuang, Quantum Computation and Quantum Information (Cambridge University Press, Cambridge, UK, 2000).
\bibitem{asm07} S. Asmussen, P. W. Glynn, Stochastic Simulation: Algorithms and Analysis (Springer-Verlag, 2007).
\bibitem{bin02} K. Binder, D. W. Heermann, Monte Carlo Simulation in Statistical Physics. An Introduction. 4th ed. (Springer, Berlin 2002).
\bibitem{bla00} Ya. M. Blanter, M. B\"uttiker, Shot noise in mesoscopic conductors.  
Phys. Rep. {\bf 336}, 1 (2000).
\bibitem{rey03} L.-H. Reydellet, P. Roche, D. C. Glattli, B. Etienne, Y. Jin, Quantum partition noise of photon-created electron-hole pairs. 
Phys. Rev. Lett. {\bf 90}, 176803 (2003). 
\bibitem{fev18} P. F\'evrier, J. Gabelli, Tunneling time probed by quantum shot noise.  
Nat. Comm. {\bf 9}, 4940 (2018). 
\bibitem{uch08} A. Uchida, K. Amano, M. Inoue, K. Hirano, S. Naito, H. Someya, I. Oowada, T. Kurashige, M. Shiki, S. Yoshimori, K. Yoshimura, P. Davis, Fast physical random bit generation with chaotic semiconductor lasers. 
Nature Photonics {\bf 2}, 728-732 (2008).
\bibitem{rei09} I. Reidler, Y. Aviad, M. Rosenbluh, I. Kanter, Ultrahigh-Speed Random Number Generation Based on a Chaotic Semiconductor Laser.
 Phys. Rev. Lett. {\bf 103}, 024102 (2009).
\bibitem{kan10} I. Kanter, Y. Aviad, I. Reidler, E. Cohen and M. Rosenbluh, An optical ultrafast random bit generator.  
Nature Photonics {\bf 4}, 58 
(2010).
\bibitem{sci15} M. Sciamanna, K.A. Shore, Physics and applications of laser diode chaos. Nature Photonics {\bf 9}, 151 (2015).
\bibitem{tan15} X. Tang, Z. M. Wu, J. G. Wu, T. Deng, J. J. Chen, L. Fan, Z. Q. Zhong, G. Q. Xia, Tbits/s physical random bit generation based on mutually coupled semiconductor laser chaotic entropy source.  
Opt. Express {\bf 23}(26), 33130 
(2015).
\bibitem{kas95}J. Kastrup, R. Klann, H. T. Grahn, K. Ploog, L. L. Bonilla, J. Gal\'an, M. Kindelan, M. Moscoso, R. Merlin,  Self-oscillations of domains in doped  GaAs-AlAs superlattices. Phys. Rev. B {\bf 52}, 13761 (1995).
\bibitem{bon02r} L.L. Bonilla, Theory of Nonlinear Charge Transport, Wave Propagation and Self-oscillations in Semiconductor Superlattices. 
J. Phys. Cond. Matter {\bf 14}, R341 
(2002).
\bibitem{bon05} L. L. Bonilla, H. T. Grahn, Nonlinear dynamics of semiconductor superlattices.
Rep. Prog. Phys. {\bf 68}, 577 
(2005).
\bibitem{BT10} L. L. Bonilla, S. W. Teitsworth, Nonlinear wave methods for charge transport (Wiley-VCH, Weinheim, 2010).
\bibitem{dre10} W. Dreyer, J. Jamnik, C. Guhlke, R. Huth, M. Gaberscek, The thermodynamic origin of hysteresis in insertion batteries. Nature Mater. {\bf 9}, 448 (2010).
\bibitem{bon15}L.L. Bonilla, A. Carpio, A. Prados, Theory of force-extension curves for modular proteins and DNA hairpins. Phys. Rev. E {\bf 91}, 052712 (2015).
\bibitem{ali13}K. Alim, G. Anselem, F. Peaudecerf, M. P. Brenner, A. Pringle, Random network peristalsis in {\em Physarum polycephalum} organizes fluid flows across an individual. Proc. Natl. Acad. Sci. {\bf 110}, 13306 (2013).
\bibitem{alm19} T. Almqvist, D. O. Winge, E. Dupont, A. Wacker, Domain formation and self-sustained oscillations in quantum cascade lasers. 
Eur. Phys. J. B {\bf 92}, 72 (2019).
\bibitem{spi19}O. Spitz, J. Wu, M. Carras, C.-W. Wong, F. Grillot, Chaotic optical power dropouts driven by low frequency bias forcing in a mid-infrared quantum cascade laser. 
Scientific Reports {\bf 9}, 4451 (2019).
\bibitem{ond21} D. E. \"Onder, A. A. S. Kalaee, D. O. Winge, A. Wacker, Chaotic Behavior of Quantum Cascade Lasers at Ignition. 
arXiv:2103.08337.
\bibitem{bom12} Yu. Bomze, R. Hey, H.T. Grahn, S.W. Teitsworth, Noise-Induced Current Switching in Semiconductor Superlattices: Observation of Nonexponential Kinetics in a High-Dimensional System. 
Phys. Rev. Lett. {\bf 109},  026801 (2012).
\bibitem{yin17}Z. Z. Yin, H. L. Song, Y. H. Zhang, M. Ruiz-Garcia, M. Carretero, L. L. Bonilla, K. Biermann, H. T. Grahn, Noise-enhanced chaos in a weakly coupled GaAs/(Al,Ga)As superlattice.  
Phys. Rev. E {\bf 95}, 012218 (2017).
\bibitem{li15} W. Li, Y. Aviad, I. Reidler, H. Song, Y. Huang, K. Biermann, Y. Zhang, H.T. Grahn, I. Kanter, Chaos synchronization in networks of semiconductor superlattices. 
Europhys. Lett. EPL {\bf 112}, 30007 (2015).
\bibitem{mom18}E. Mompo, M. Ruiz-Garcia, M. Carretero, H. T. Grahn, Y. H. Zhang, L. L. Bonilla, Coherence Resonance and Stochastic Resonance in an Excitable Semiconductor Superlattice.  
Phys. Rev. Lett. {\bf 121}, 086805 (2018).
\bibitem{sha18} Z. Z. Shao, Z. Z. Yin, H. L. Song, W. Liu, X. J. Li, J. Zhu, K. Biermann, L. L. Bonilla, H. T. Grahn, Y. H. Zhang, Fast Detection of a Weak Signal by a Stochastic Resonance Induced by a Coherence Resonance in an Excitable GaAs/Al$_{0.45}$Ga$_{0.55}$As Superlattice. 
Phys. Rev. Lett. {\bf 121}, 086806 (2018).
\bibitem{alv14} M. Alvaro, M. Carretero, L.L. Bonilla, Noise enhanced spontaneous chaos in semiconductor superlattices at room temperature. 
Europhys. Lett. (EPL) {\bf 107}, 37002 (2014).
\bibitem{bon17} L.L. Bonilla, M. Alvaro, M. Carretero, Chaos-based true random number generators. 
J. Math. Ind. {\bf 7}, 1 (2017).
\bibitem{rui17} M. Ruiz-Garcia, J. Essen, M. Carretero, L. L. Bonilla, B. Birnir, Enhancing Chaotic Behavior at room temperature in GaAs/(Al,Ga)As Superlattices.  
Phys. Rev. B {\bf 95}, 085204 (2017).
\bibitem{ess18} J. Essen, M. Ruiz-Garcia, I. Jenkins, M. Carretero, L. L. Bonilla, B. Birnir, Parameter dependence of high-frequency nonlinear oscillations and intrinsic chaos in short GaAs/(Al,Ga)As superlattices. 
Chaos {\bf 28}, 043107 (2018). 
\bibitem{ama02} A. Amann, J. Schlesner, A. Wacker, E. Sch\"oll, Chaotic front dynamics in semiconductor superlattices. 
Phys. Rev. B {\bf 65}, 193313 (2002).
\bibitem{ama03}A. Amann, K. Peters, U. Parlitz, A. Wacker, E. Sch\"oll, A hybrid model for chaotic front dynamics. 
Phys. Rev. Lett. {\bf 91}, 066601 (2003).
\bibitem{bul95} O. M. Bulashenko, L. L. Bonilla,  Chaos in resonant-tunneling superlattices. 
Phys. Rev. B {\bf 52}, 7849 
(1995).
\bibitem{luo98}K. J. Luo, H. T. Grahn, K. H. Ploog, L. L. Bonilla,  Explosive bifurcation to chaos in weakly-coupled semiconductor superlattices.
Phys. Rev. Lett. {\bf 81}, 1290 
 (1998).
\bibitem{wac02} A. Wacker, Semiconductor superlattices: A model system for nonlinear transport.  
Phys. Rep. {\bf 357}, 1 
(2002).
\bibitem{agu97} R. Aguado, G. Platero, M. Moscoso, L. L. Bonilla, Microscopic  Model for Sequential Tunneling in Semiconductor Multiple Quantum Wells.  
Phys. Rev. B {\bf 55}, R16053 
 (1997).
\bibitem{bon00} L. L. Bonilla, G. Platero, D. S\'anchez, Microscopic derivation of transport coefficients and boundary conditions in discrete drift-diffusion models of weakly coupled superlattices. 
Phys. Rev. B {\bf 62}, 2786 
 (2000).
 \bibitem{gol87a}
V. J. Goldman, D. C. Tsui, J. E. Cunningham, Resonant tunneling  in magnetic fields: evidence for space-charge buildup. Phys. Rev. B {\bf 35}, 9387 (1987).
\bibitem{suppl} See Supplemental Material for expressions of tunneling currents, evolution of the current and the density plot of the electric field profile for periodic and chaotic attractors, and quantification of the effect that randomness in well width has on the chaotic attractors.
\bibitem{ama01} A. Amann, A. Wacker, L. L. Bonilla, E. Sch\"oll, Dynamic scenarios of multistable switching in semiconductor superlattices. 
Phys. Rev. E {\bf 63}, 066207 (2001). 
\bibitem{bon06} L. L. Bonilla, R. Escobedo, G. Dell'Acqua, Voltage switching and domain relocation in semiconductor superlattices. 
Phys. Rev. B {\bf 73}, 115341 (2006).

\bibitem{gra95} H. T. Grahn (editor), Semiconductor superlattices. Growth and electronic properties (World Scientific, Singapore 1995).

\bibitem{das12} P. Dasmahapatra, J. Sexton, M. Missous, C. Shao, and M. J. Kelly, Thickness control of molecular beam epitaxy grown layers at the 0.01-0.1 monolayer level. Semicond. Sci. Technol. {\bf 27}, 085007 (2012).

\bibitem{mis15} M. Missous, M. J. Kelly and J. Sexton, Extremely Uniform Tunnel Barriers for Low-Cost Device Manufacture. IEEE Electron Device Letters {\bf 36}, 543 (2015).

\bibitem{zho02} C. Zhou and J. Kurths, Noise-induced Phase Synchronization and Synchronization Transitions in Chaotic Oscillators. Phys. Rev. Lett. {\bf 88}, 230602 (2002).

\bibitem{keu17} L. Keuninckx, M. C. Soriano, I. Fischer, C. R. Mirasso, R. M. Nguimdo, G. Van der Sande,  Encryption key distribution via chaos synchronization. 
Sci. Rep. {\bf 7}, 43428 (2017).


\bibitem{liu18} W. Liu, Z. Yin, X. Chen, Z. Peng, H. Song, P. Liu, X. Tong, Y. Zhang, A secret key distribution technique based on semiconductor superlattice chaos devices.  
Science Bulletin {\bf 63}, 1034 
(2018).
\end{thebibliography}
\end{document}